\documentclass[preprint,showpacs,preprintnumbers,amsmath,amssymb]{revtex4}

\usepackage{graphicx}
\usepackage{dcolumn}
\usepackage{bm}

\begin{document}

\preprint{APS/123-QED}

\title{Refocussing off-resonant spin-1/2 evolution using spinor behavior}

\author{Jonathan Baugh}
\affiliation{%
Institute for Quantum Computing and Department of Chemistry\\
University of Waterloo, Waterloo, Ontario, N2L 3G1
}%

\date{\today}

\begin{abstract}
A systematic method for constructing increasingly precise pulse sequences to refocus spin evolution in the presence of large resonance offset is presented. Error cancellation relies, in part, on the spinor transformation property that yields a phase $(-1)^k$ for a rotation of $2k\pi$ (integer $k$). The sequences only require the ability to perform $\pi$ pulses about two opposite axes, lending simplicity to experimental implementation. Very high fidelities with the identity operator can be achieved for resonance field offsets comparable to the resonant field amplitude, particularly when using appropriate composite $\pi$ rotations as base pulses. The sequences possess good robustness to systematic amplitude errors, and for certain choice of base pulse, a systematic under-rotation allows the bandwidth of high-fidelity refocussing to be extended to offsets larger than resonant field amplitude. 
\end{abstract}

\maketitle

\section{\label{sec:level1}Introduction}
Utilizing spins as quantum bits (qubits) for solid-state quantum information processing (QIP) \cite{Nielsen2000a, Loss1998} poses many practical challenges, but particularly acute are those concerning the manipulation and coherence of spin states given limited controls and a real environment. The problem of controlling a single spin with an unknown, static resonance field offset is often encountered in magnetic resonance, and is pertinent to recent experiments on single electron spin resonance (SESR) in semiconductor quantum dots containing lattice nuclear spins \cite{Koppens2006a,Nowack2007a,Pioro-Ladriere2008a}. In this paper, we shall focus on the special problem of refocussing spin evolution (i.e. generating a pulse sequence to approximate the identity operator) under resonance field offsets comparable to the control field amplitude. To date, SESR experiments using applied microwave fields \cite{Koppens2006a} or electric-dipole spin resonance \cite{Nowack2007a,Pioro-Ladriere2008a} have operated in a regime in which the distribution of random, quasi-static nuclear fields \footnote{For external magnetic field $\vec{B}=B\hat{z}$ much larger than the random nuclear field $\vec{B}_N$, the transverse components of $\vec{B}_N$ do not commute with $\vec{B}$ and can be ignored to very good approximation.} gives rise to unknown resonance frequency offsets larger than the available Rabi frequencies. Recent experiments, however, suggest that technological improvements could make significantly faster spin rotations available \cite{Pioro-Ladriere2008a}, so that control amplitude and offset may be comparable. The ability to refocus spin evolution with high fidelities in this setting would provide a means for precise measurement of decoherence times and for preserving qubit coherence over measurement ensembles. A robust method that is simple to implement experimentally would be especially desirable. \\
\indent One approach to this problem is to focus on improving the fidelity of a single $\pi$ rotation by designing composite pulses that cancel errors to successively higher orders, at the expense of exponential growth in pulse length with respect to leading-order error \cite{Levitt1986, Brown2004}. The optimal pulse sequence for a desired period $T$ would then be $P-\tau-P-\tau$, where $2(\tau+t_P)=T$ and $P$ represents a composite $\pi$-pulse of duration $t_P$. Optimal performance would be expected for $t_P=T/2$, i.e. use of the highest fidelity (longest) composite pulse. Conceptually this is an attractive approach, but so far it has proven difficult to construct very high order composite pulses to refocus off-resonance errors \cite{Alway2007,Brown2004}, as the finite excitation bandwidth for a given resonant field amplitude (and control bandwidth) ultimately limits the achievable accuracy for arbitrary offset. While these fundamental limits apply to any approach, other methods have been developed to generate robust decoupling sequences from limited controls, such as Eulerian cycles on a group \cite{Viola2002}, and concatenated dynamical decoupling sequences with imperfect pulses \cite{Khodjasteh2007}. This paper demonstrates an alternative method of approximating the identity operator in the special case of large resonance offsets, that allows errors to be cancelled to arbitrary order in principle, and that should lend itself well to experimental implementation. We assume a base $\pi$-pulse $P$ with some fixed fidelity profile versus resonance offset, and then construct sequences using $P$ and its phase-reversed twin pulse $\bar{P}$ (i.e. a rotation about the opposite axis in the Bloch sphere). Similar to the case of a single composite pulse, it is evident that a constant amplitude (windowless) sequence of pulses can make maximum use of the available excitation and control bandwidths. Given a $2\pi$ pulse duration $1/\nu_1$ ($\nu_1$ is the resonant field amplitude) that is an integral fraction of $T$, we look for sequences of $N=2\nu_1T$ $\pi$-pulses that will best approximate the identity operator. Our algorithm for constructing these error-compensating sequences relies explicitly on the transformation property of spinors that an ideal rotation of $2k\pi$ generates an evolution operator $(-1)^k\mathbf{1}$, where $\mathbf{1}$ is the identity matrix and $k\geq 0$ is an integer. As longer sequences are constructed, higher order errors due to resonance offset are canceled, so that extremely high fidelities are reached for $\epsilon<<1$, and refocussing bandwidths can be pushed out to $\epsilon\sim1$, limited ultimately by the bandwidth of control (a property of the chosen base pulse $P$). Since the sequences are built up from only two base pulses, $P$ and $\bar{P}$, experimental implementation is simple compared to high-order composite pulses or numerically optimized control sequences. We also show that these sequences can be applied in windowed mode (i.e. with interspersed, equal delays) without appreciably degrading the fidelity. A benefit of multi-pulse refocussing relevant to QIP applications is suppression of homogeneous dephasing due to spectral diffusion of a coupled bath, when the bath correlation time is much longer than the spin nutation period. Indeed, this is the expected working regime for an electron spin qubit coupled to a nuclear spin bath in a quantum dot \cite{Coish2009a}. However, we set aside the detailed analysis of these sequences under dynamic errors for future work. 
\subsection{\label{sec:level2} Brief review of composite pulses}
Treating resonance offset errors in the context of average Hamiltonian theory \cite{Haeberlen1976a, Tycko1983}, we can write the time-dependent propagator of a pulse as $U(\tau) = U_{ideal}(\tau) U_{error}(\tau)$ where $\tau$ is the pulse length and $U_{error}(t) = \mathcal{T}e^{-i \int_0^t \tilde{V}(t') dt'}$. The interaction frame error Hamiltonian is $\tilde{V}(t) =  U^{-1}_{ideal}(t)(\pi\epsilon\nu_1 \sigma_z) U_{ideal}(t)$, where $\epsilon$ is the fractional resonance offset with respect to resonant field amplitude $\nu_1$, and $\sigma_z$ the $\hat{z}$ Pauli matrix. Tycko \cite{Tycko1983} has designed wideband composite pulses by numerically minimizing the lowest order average Hamiltonian terms in the Magnus expansion for $U_{error}(\tau)$, a conceptually elegant but computationally tedious method. Other methods have been used to construct composite rotations robust to both off-resonant and control field amplitude errors, such as the construction of broadband pulses by Wimperis \cite{Wimperis1994}, and the quaternion-based formalism of Cummins and Jones \cite{Cummins2003,Xiao2006}. Fundamental concepts and properties of composite pulses are laid out in a review by Levitt \cite{Levitt1986}. Brown et al. devised formal methods for generating arbitrarily accurate sequences (for arbitrary single qubit rotations) that compensate for general errors \cite{Brown2004}, and provided specific constructions for pulse length errors but not for off-resonant errors. More recent work on implementation of robust state transformations and unitary operators has been carried out in the context of optimal control theory, e.g. use of the GRAPE method \cite{Khaneja2005, Kobzar2004, Ryan2008a} for numerical optimization of pulse waveforms. Given some set of pulses with fixed precision, we are faced with the challenge of how to design sequences of many pulses that will fully utilize the control bandwidth so that errors do not accumulate. Indeed, an optimal refocussing sequence should always be non-repeating, and tailored to the desired refocussing period.  The sequences derived below use symmetry and the spinor transformation property to strategically cancel errors as longer sequences are constructed. 
\section{Sequences}
\subsection{\label{sec:level3} Definitions}
In the following we will represent 2$\times$2 unitary matrices as $U = a \mathbf{1} + i\vec{b} \cdot \vec{\sigma}$, where $(\sigma_{x}, \sigma_{y}, \sigma_{z})$ are the Pauli matrices and $a^2+|\vec{b}|^2=1$. Turning on a resonant field of strength $\omega_1$ along the $\hat{x}$ direction produces a unitary: 
\begin{align}
U_x(\theta, \epsilon) &= e^{i \int_0^{\theta/\omega{1}} \frac{\omega_1}{2} (\sigma_x + \epsilon \sigma_z)dt}\nonumber \\
&= a(\theta,\epsilon)\mathbf{1} + i b_x(\theta,\epsilon) \sigma_x + i b_z(\theta,\epsilon) \sigma_z
\end{align}
where $\epsilon \omega_1$ is the resonance offset, $b_z \sim \mathcal{O}(\epsilon)$ and $b_x\approx \mathrm{sin}(\theta/2)+\mathcal{O}(\epsilon^2)$. We denote the unitary corresponding to a phase-reversed resonant field as $U_{\bar{x}} = a\mathbf{1} - i b_x \sigma_x + i b_z \sigma_z$. We shall be interested in general unitaries of two forms, approximate rotations by $\pi$ and $2k\pi$ ($k=0,1,2,\dots$), respectively:
\begin{align}
U_\pi : \hspace{5mm}&a = g_1(\epsilon^2)\nonumber \\
             &\vec{b} = (\cos{\phi} + g_x(\epsilon),\sin{\phi} + g_y(\epsilon), g_z(\epsilon))\\
U_\mathbf{1} : \hspace{5mm}& a = (-1)^k + g'_1(\epsilon^2) \nonumber\\
              &\vec{b} =(g'_x(\epsilon), g'_y(\epsilon), g'_z(\epsilon))
\end{align}
where the functions denoted $g,g'$ are the \emph{error} terms that, in general, can be at any order in $\epsilon$ (or $\epsilon^2$). The rotation axis for $U_\pi$ is $\hat{x}\cos{\phi} + \hat{y}\sin{\phi}$ for arbitrary $\phi$.\\
\indent Observe that the combined unitary $U_\alpha U_{\bar{\alpha}}$, which we term \emph{antisymmetric}, has leading-order $\sigma_z$ coefficient $b'_z = 2 a b_z$, and transverse coefficients $b'_y = 2 b_x b_z$, $b'_x = 2 b_y b_z$. The $\sigma_z$ term is the same as for the symmetric version $U_\alpha U_\alpha$, whereas the transverse terms are smaller by the factor $b_z$. Hence, the transverse error for an approximate identity sequence is significantly reduced by using the antisymmetric structure, but only in proportion to the existing $\sigma_z$ error. Therefore the total error of an identity sequence ultimately depends on how far the $\sigma_z$ error can be reduced by construction of the sequence; it can be guaranteed that the $\sigma_z$ error will be dominant for $\epsilon<<1$ at the expense of doubling the sequence length. We also note that a \emph{symmetric-antisymmetric} (SA) sequence of the form $U_\alpha U_{\bar{\alpha}} U_{\bar{\alpha}} U_\alpha $ reduces the transverse error by an additional factor $b_z$ so that it is $\sim\mathcal{O}(b_{x,y} b^2_z)$.\\
\begin{table*}[!ht]
\begin{center}
  \begin{tabular}{ |l || c | c | c | c |}
    \hline
     \large{\textbf{Sequence}} &  $\mathbf{\mathcal{O}(\delta_x)}$ & $\mathbf{\mathcal{O}(\delta_y})$ &$\mathbf{\mathcal{O}(\delta_z)}$ & $|\epsilon_{max}|, f=0.99$ \small{\{simple, 3-pulse, 7-pulse\}}\\ \hline
 \scriptsize{`2'=$PP$} & $b_x a$ & $ a\sim \epsilon^2$ &$a b_z\sim \epsilon^3$ & 0.25\hspace{1mm},\hspace{1mm} 0.47\hspace{1mm},\hspace{1mm} 0.72\\ \hline
  \scriptsize{`4'=$PP\bar{P}\bar{P}$} & $b_x a^2b_z$ & $ a^2b_z\sim \epsilon^5$ &$a b_z\sim \epsilon^3$ & 0.32\hspace{1mm},\hspace{1mm} 0.54\hspace{1mm},\hspace{1mm} 0.74\\ \hline
    \scriptsize{`8'=$PP\bar{P}\bar{P}\bar{P}\bar{P}PP$} & $a^3b^2_z$ &$a^3b^2_z\sim \epsilon^8$  & $a b_z \sim \epsilon^3$& 0.25\hspace{1mm},\hspace{1mm} 0.49\hspace{1mm},\hspace{1mm} 0.72\\ \hline
      \scriptsize{`16'=$A\bar{A}$} &$a^3b^4_z$  &$a^3b^4_z\sim \epsilon^{10}$  &$ab^3_z \sim \epsilon^5$ & 0.43\hspace{1mm},\hspace{1mm} 0.77\hspace{1mm},\hspace{1mm} 0.79\\ \hline
        \scriptsize{`32'=$A\bar{A}\bar{A}A$} & $a^3b^8_z$ &$a^3b^8_z\sim \epsilon^{14}$  &$ab^3_z \sim \epsilon^5$  &0.36\hspace{1mm},\hspace{1mm} 0.60\hspace{1mm},\hspace{1mm} 0.76 \\ \hline
          \scriptsize{`64'=$C\bar{C}$} & $a^2b^9_z$ &$a^2b^9_z\sim \epsilon^{13}$  & $ab^5_z \sim \epsilon^7$& 0.73\hspace{1mm},\hspace{1mm} 0.80\hspace{1mm},\hspace{1mm} 0.87 \\ \hline
            \scriptsize{`128'=$C\bar{C}\bar{C}C$} &$a^3b^{14}_z$  &$a^3b^{14}_z\sim \epsilon^{20}$  &$ab^5_z \sim \epsilon^7$ & 0.45\hspace{1mm},\hspace{1mm} 0.79\hspace{1mm},\hspace{1mm} 0.86\\ \hline
             \scriptsize{`256'=$F\bar{F}$} &$a^3b^{12}_z$  &$a^3b^{12}_z \sim \epsilon^{18}$  &$ab^7_z \sim \epsilon^9$ & 0.72\hspace{1mm},\hspace{1mm} 0.79\hspace{1mm},\hspace{1mm} 0.86\\ \hline
  \end{tabular}
\end{center}
\caption{\label{tab:numerical} Characteristics of successively larger `optimized' approximate identity sequences, where `N' indicates the number of base $\pi_y$($\pi_{\bar{y}}$) pulses $P$($\bar{P}$). Residual orders of error terms $\vec{\delta}$ for $\sigma_x$, $\sigma_y$ and $\sigma_z$ operators are given in the notation of equation $2$. The maximum fractional offsets $|\epsilon_{max}|$ for $10^{-2}$ infidelity with the identity operator are shown using a simple $\pi$-pulse, a 3-pulse composite $\pi$ \cite{Levitt1986}, and a 7-pulse composite $\pi$ \cite{Tycko1983} as base pulses. Sequences $A$, $C$ and $F$ are defined in the text, section~\ref{silly}.}
\end{table*}
\begin{figure*}[!h]
 \includegraphics[width=54mm]{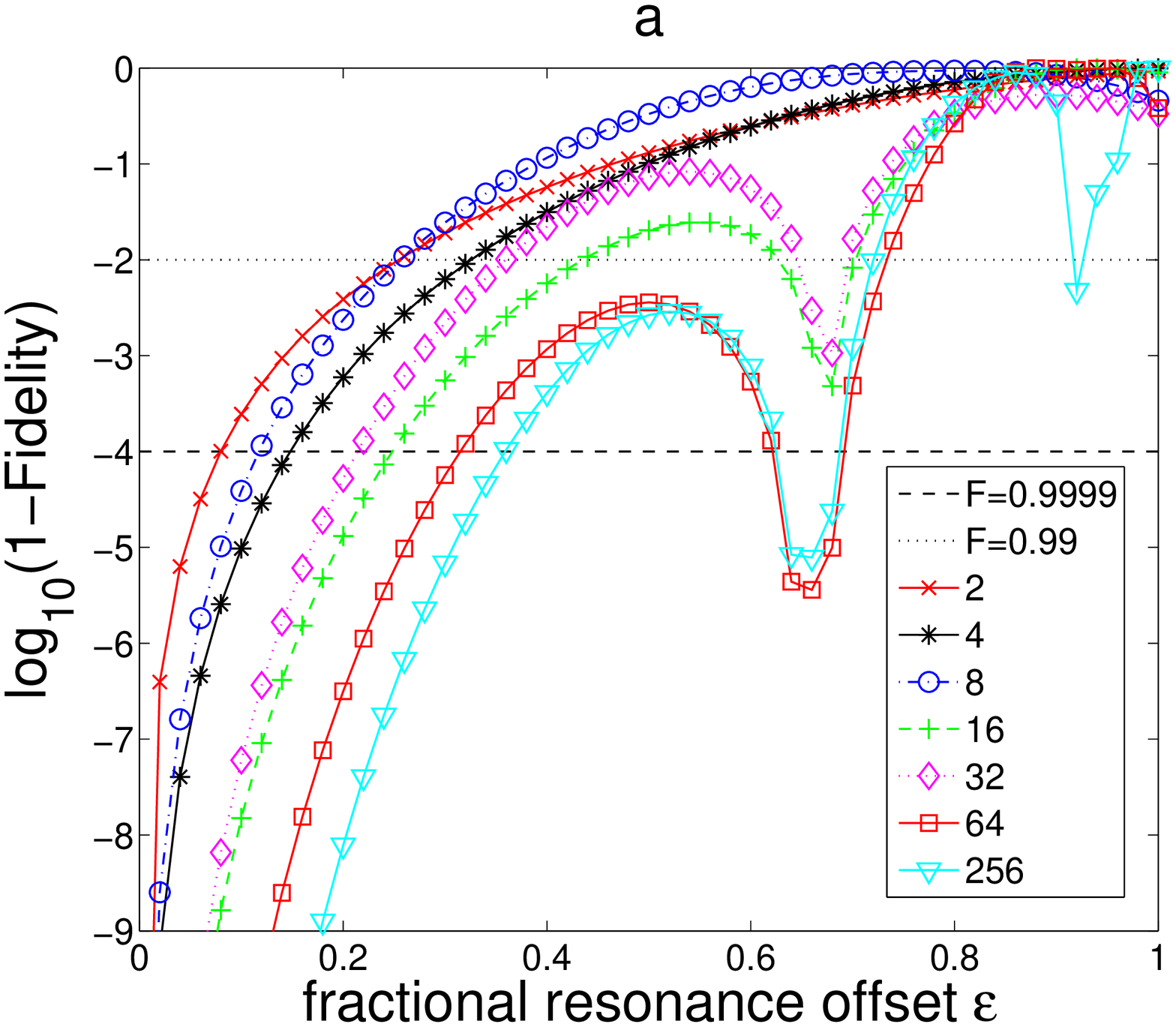}
  \includegraphics[width=54mm]{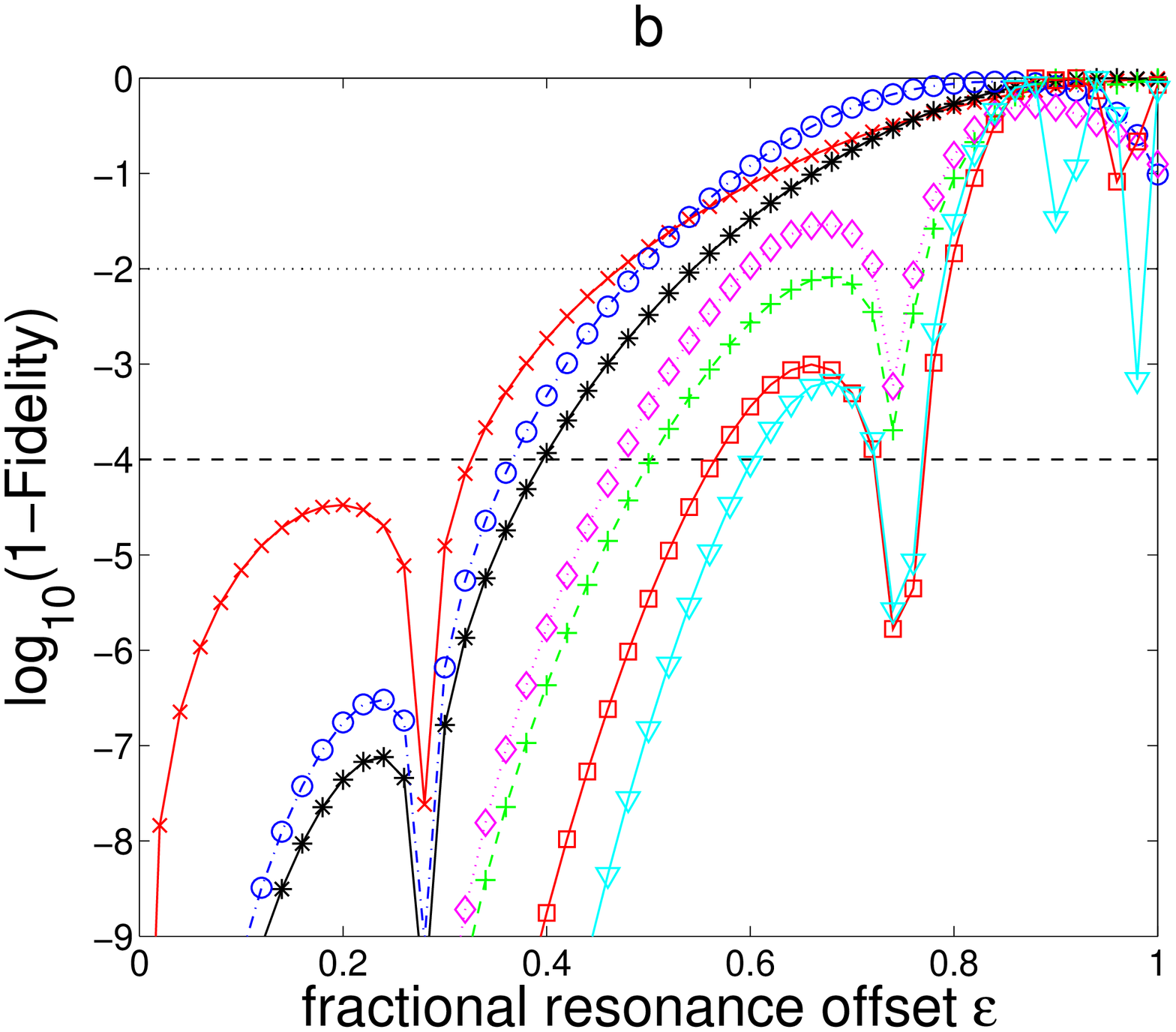}
   \includegraphics[width=54mm]{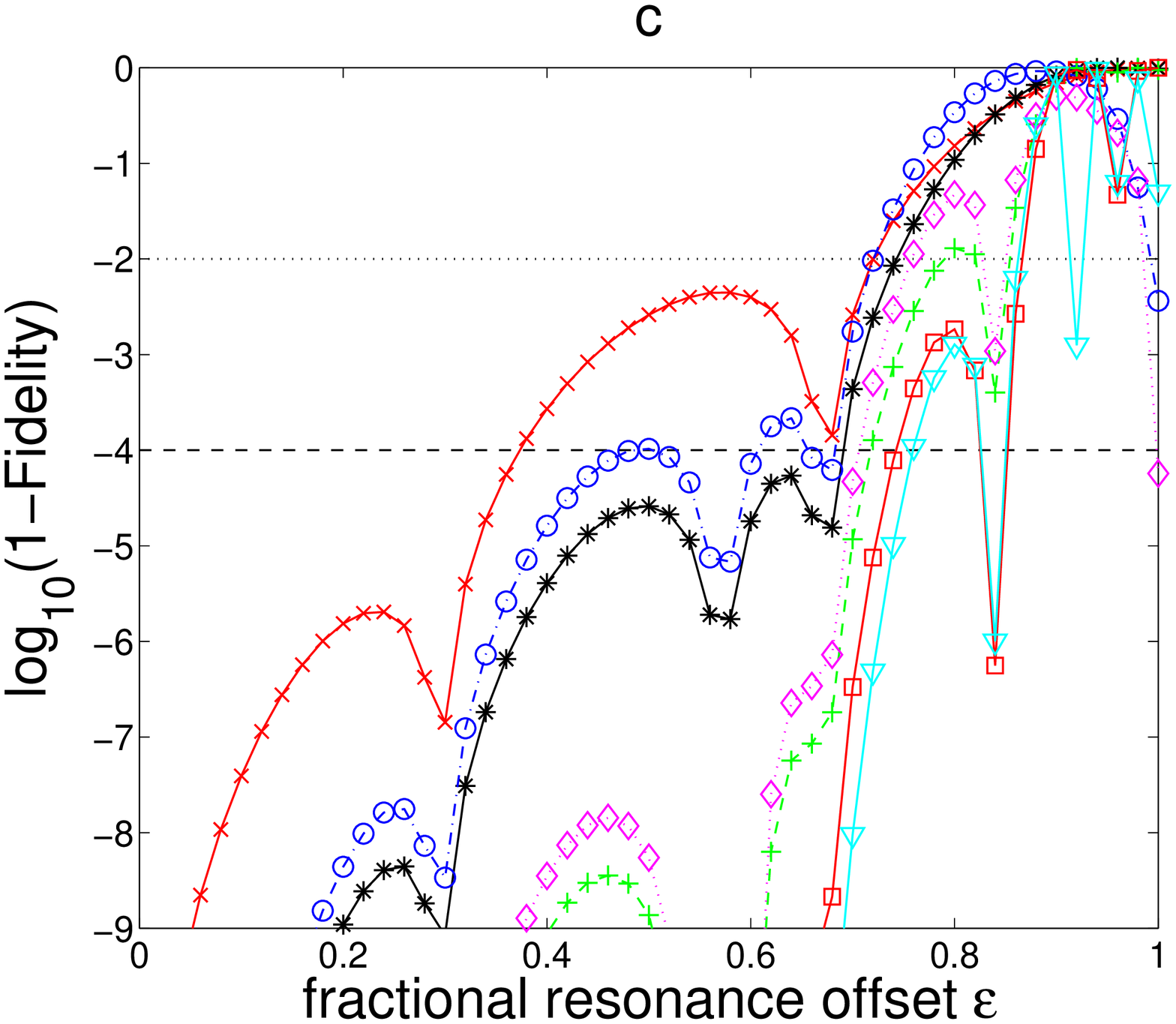}
\caption{\label{fig:fig1} Numerical results for logarithmic infidelity versus fractional resonance offset $\epsilon$ for the (windowless) pulse sequences listed in Table 1. $\epsilon=1$ corresponds to the resonance offset equal to the control field amplitude. Horizontal dotted lines indicate fidelities of $0.99$ and $0.9999$. Since these are windowless sequences, the total time for each sequence is the base pulse duration $t_P \times N$, where $N$ is the number of base pulses. \textbf{(a)} Base pulse $P$ is a simple $\pi$ pulse. \textbf{(b)} $P$ is the three-pulse composite $\pi$ rotation $(\pi/2)_x(3\pi/2)_y(\pi/2)_x$ \cite{Levitt1986}.  \textbf{(c)} $P$ is a 7-pulse composite $\pi$ rotation \cite{Tycko1983}.}
\end{figure*}
\subsection{\label{silly} Construction algorithm}
\begin{enumerate}
\item Find two sequences $S_1, S_2$ that have equal leading order $\sigma_z$ error: $S_1$ to generate a $2k\pi$ ($k$ even) rotation and $S_2$ to generate a $2k'\pi$ ($k'$ odd) rotation; 
\item Note that $S_1\bar{S_1}\approx \mathbf{1}+...+2\xi_z \sigma_z $ and $S_2\bar{S_2}\approx \mathbf{1}+...-2\xi_z \sigma_z $, where $\xi_z$ is the leading order error of both $S_1$ and $S_2$, and the $-\xi_z$ in the second expression is due to the spinor rotation property; then form the combined sequence $A=S_1\bar{S_1}S_2\bar{S_2}$ which has a reduced $\sigma_z$ error $\xi'_z\sim\mathcal{O}(\epsilon^p)$ due to cancellation of $\xi_z\sim\mathcal{O}(\epsilon^q)$ terms, where $p>q$. 
\item From $A$ we can form two useful sequences, antisymmetric $A\bar{A}$ and SA sequence $A\bar{A}\bar{A}A$. 
\item For the next larger sequence, note that $A$ is a $0 \pi$ rotation, and find a sequence $B$ of the same length that generates a $2k'\pi$ ($k'$ odd) rotation and has the same leading-order $\sigma_z$ error. (For $k'=1$, $B$ can be found that only differs from $A$ by the phase reversal of one particular base pulse); 
\item label $S_1=A$ and $S_2=B$, jump to step 2. 
\end{enumerate}
For example, labelling $P = U_\pi$ and $\bar{P} = U_{\bar{\pi}}$, the two elementary sequences $PP$ and $P\bar{P}$ can be used for step 1: 
\begin{align}
S_1=P\bar{P}& \approx \mathbf{1}-i 2 b_z \sigma_x+ i 2 b_x b_z \sigma_y+i 2 a b_z \sigma_z \label{eq:1}\\
S_2=PP& \approx -\mathbf{1}+i 2 a b_x \sigma_x+ i 2 a \sigma_y+i 2 a b_z \sigma_z \label{eq:2} 
\end{align}
Note we have taken $b_y=1+\mathcal{O}(\epsilon^2)$ (i.e. $P$ is an approximate $\pi$ pulse along $\sigma_y$), without loss of generality. The leading coefficient of $\mathbf{1}$ in the $PP$ sequence is $-1$ owing to transformation of a spinor under $2\pi$ rotation. We have:
\begin{align}
S_1\bar{S_1} \approx \mathbf{1}+...+i 4 a b_z \sigma_z \label{eq:3} \\
S_2\bar{S_2} \approx \mathbf{1}+...-i 4 a b_z \sigma_z \label{eq:4}
\end{align}
so that
\begin{align}
A=S_1\bar{S_1}S_2\bar{S_2} \approx \mathbf{1}+...+\mathcal{O}(a^3 b_z+a b^3_z) \sigma_z \label{eq:5}
\end{align}
The 16-pulse antisymmetric sequence $A\bar{A}$, where $\bar{A}=\bar{S_1}S_1\bar{S_2}S_2$, has $\sigma_z$ error $\sim\mathcal{O}(a^3 b_z+a b^3_z)\sim\mathcal{O}(\epsilon^5)$, and transverse ($\sigma_x$, $\sigma_y$) errors $\sim\mathcal{O}(a^2 b^3_z)\sim\mathcal{O}(\epsilon^7)$. \\
\indent Next, we take $A$ as the $0\pi$ rotation and find another sequence $B$ for a $2\pi$ rotation with the same $\sigma_z$ error. It can be verified that the sequence $\bar{P}\bar{P}\bar{P}PPP\bar{P}\bar{P}$ (i.e. the same as $A$ but with the first pulse reversed in phase) satisfies this condition. We can then construct $C=A\bar{A}B\bar{B}$ which has leading-order $\sigma_z$ error reduced to $\sim\mathcal{O}(a b^5_z)\sim\mathcal{O}(\epsilon^7)$, yielding the useful 64- and 128-pulse sequences $C\bar{C}$ and $C\bar{C}\bar{C}C$. The leading transverse errors of $C\bar{C}$ are $\sim\mathcal{O}(a^2 b^9_z)\sim\mathcal{O}(\epsilon^{13})$. For the next iteration, a $2\pi$ pulse that complements $C$ can be found, for example, $D=A\bar{A}A\bar{B}$, yielding a 256-pulse sequence $F\bar{F}$ where $F=C\bar{C}D\bar{D}$. Hence, the process can be iterated to any final sequence length of size $2^{4+n}$ base pulses, where $n$ is a positive integer. Since $\sigma_z$ error cannot be reduced for sequences smaller than 8 pulses (relative to the elementary sequence $PP$), the `optimal' sequences for 4 and 8 pulses are simply the antisymmetric and SA sequences $PP\bar{P}\bar{P}$ and $PP\bar{P}\bar{P}\bar{P}\bar{P}PP$, as listed in Table 1.
\begin{figure*}[!h]
 \includegraphics[width=54mm]{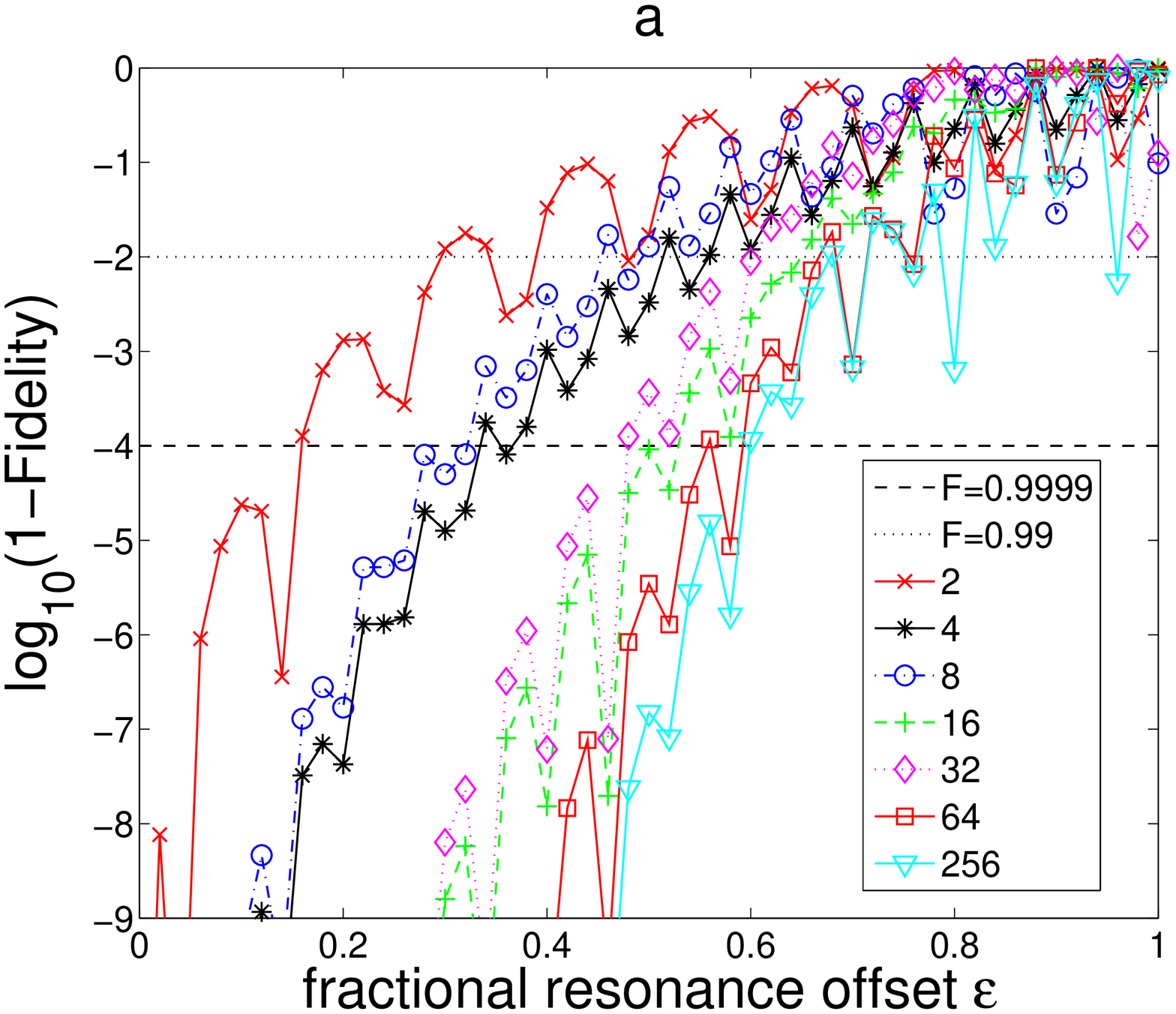}
  \includegraphics[width=54mm]{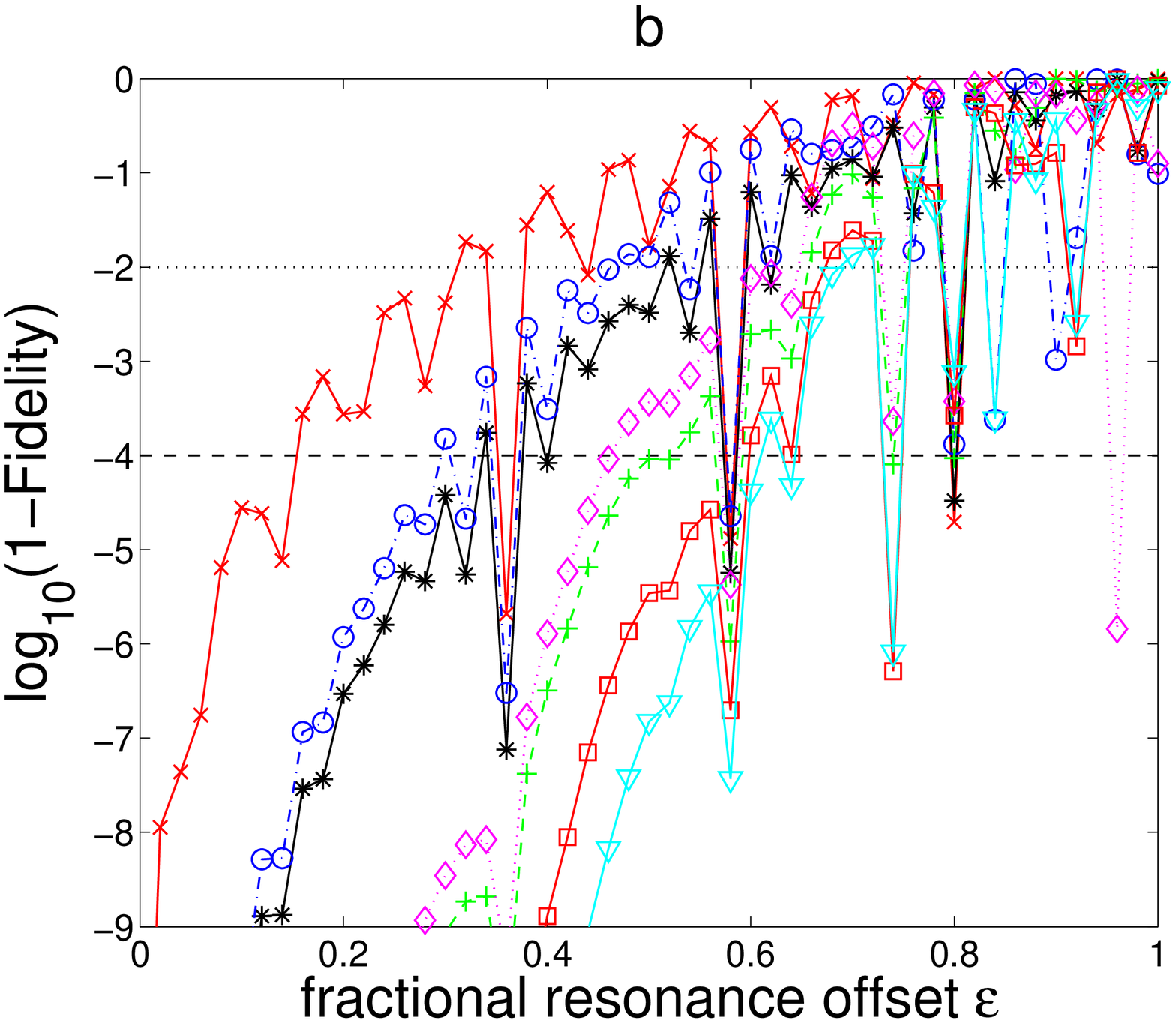}
  \includegraphics[width=54mm]{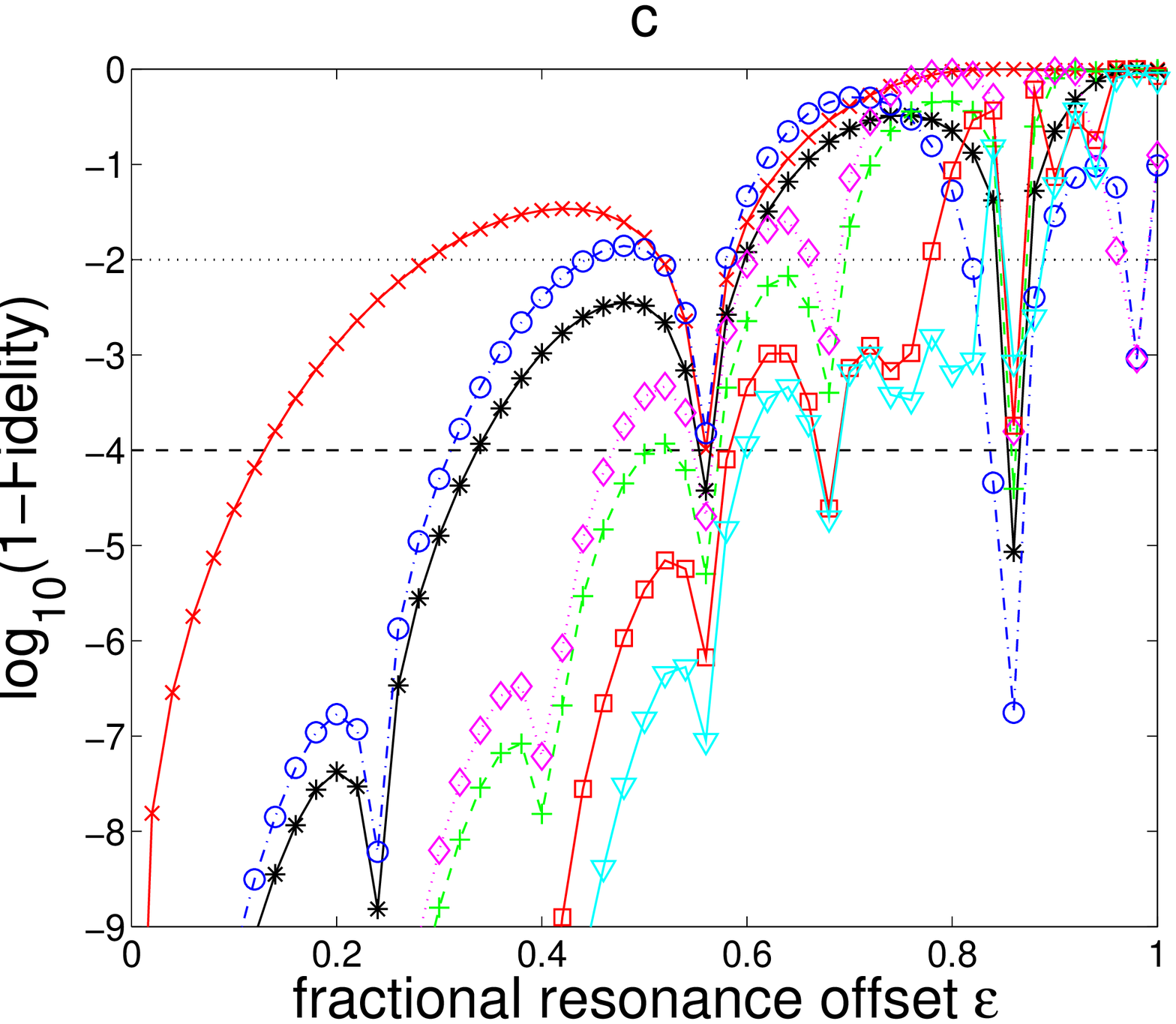}
\caption{\label{fig:fig2} Numerical results for windowed sequences, using the 3-pulse composite $\pi$ rotation. \textbf{(a)} Each base pulses is preceded by a delay period $\Delta t = 8/\nu_1$, where $\nu_1$ is the control field amplitude in Hz. \textbf{(b)} $\Delta t = 36/\nu_1$. \textbf{(c)} $\Delta t = 48/\nu_1$. Delays give rise to small amplitude oscillations of fidelity versus offset, but do not  appreciably degrade the overall sequence performance, for any delay.}
\end{figure*}
\subsection{\label{sec:level4} Performance: numerical simulation}
\indent Figure~\ref{fig:fig1} displays numerical results for the sequence fidelity with respect to the identity operator, defined as $f(\epsilon)=|Tr(U^{\dag}_{\epsilon}\mathbf{1})|^2/4$ for a sequence that generates an offset-dependent unitary $U_{\epsilon}$. Three types of base pulses are used: a simple $\pi$ pulse, and two types of composite $\pi$ rotations designed to compensate off-resonance errors. These are the 3-pulse composite rotation $(\pi/2)_x(3\pi/2)_y(\pi/2)_x$ due to Levitt  \cite{Levitt1986}, and the 7-pulse composite rotation due to Tycko \cite{Tycko1983}. Clearly, the overall performance of our sequences depends strongly on the leading-order error terms of the chosen base pulse. For both simple and composite $\pi$ pulses $P$ we generally have $a\sim\mathcal{O}(\epsilon^2)$ and $b_z\sim\mathcal{O}(\epsilon)$, and for $PP$, $a'\sim\mathcal{O}(\epsilon^4)$ and $b'_z\sim\mathcal{O}(\epsilon^3)$. For the simple pulse, all coefficients are of order unity, whereas for composite pulses they can be much smaller. For the Tycko composite pulse, $a\approx 10^{-3}\epsilon^2$ and $b_z\approx 10^{-2}\epsilon$, and for $PP$, $b'_z\sim a b_z \approx 10^{-5}\epsilon^3$. Since each final sequence is either antisymmetric or SA, we are guaranteed that the residual error will be dominated by the $\sigma_z$ error for $\epsilon<<1$. \\
\indent Figure~\ref{fig:fig1} shows that the range of offsets over which very high fidelities can be achieved improves significantly with sequence size. In particular, for the simple pulse and three-pulse cases, the `bandwidth of fidelity' at the $f>0.9999$ range improves progressively for every sequence of length $2^m$, for $m$ even. With the exception of $m=2$, these are the antisymmetric sequences for which $\sigma_z$ error is reduced compared to the previous sequence (see Table 1). Slight losses of fidelity are seen for the AS sequences with $m$ odd, as they are constructed to further suppress transverse error, but allow the dominant $\sigma_z$ error to double. For the 7-pulse Tycko sequence, the off-resonance compensation of the base pulse is particularly good, and all the non-trival sequences ($m\geq 2$) result in infidelities $\sim\mathcal{O}(10^{-4})$ to at least $|\epsilon|\approx0.67$. The improvements with sequence size at large offsets are more modest for this base pulse, for reasons discussed below. On the other hand, the longer sequences ($m\geq4$) yield substantial improvements in precision (to error $\sim\mathcal{O}(10^{-8}-10^{-9})$) for offsets up to $|\epsilon|\approx0.6-0.7$. \\
\indent For infidelity $\leq 10^{-2}$, the optimal sequence length for all three base pulse types is the 64-pulse sequence; from Table 1, we see that maximum offsets $|\epsilon_{max}|$ of 0.73, 0.80, and 0.87 can be tolerated for the simple, 3-pulse, and 7-pulse base pulses, respectively. For larger sequences, although error orders can continue to be reduced, the coefficients of the leading-order terms grow linearly with sequence size\footnote{Since the $\sigma_z$ error is dominant, upon doubling a sequence we have $(\mathbf{1}+...+h(\epsilon)\sigma_z)(\mathbf{1}+...+h'(\epsilon)\sigma_z)\approx \mathbf{1}+...+(h+h')\sigma_z$.} so that increasingly small gains in $|\epsilon_{max}| \sim 1$ are obtained. From Table 1, we see that the \emph{order} of $\sigma_z$ error is $\mathcal{O}(\epsilon^{m})$ for sequences of length $2^m$ ($m$ odd), i.e. a logarithmic increase with sequence size. Transverse error orders also increase logarithmically, but at a much faster rate ($\mathcal{O}(\epsilon^{3m-1})$ for $m$ odd). However, the linear increase in magnitude of error \emph{coefficients} severely limit the gains at large $|\epsilon|$ for sequences larger than an optimal size ($\sim 64$ pulses). It is reasonable to expect these limitations from the perspective of control bandwidth; extending refocusing to larger offsets would necessarily require either larger pulse amplitude (larger excitation bandwidth) or larger control bandwidth, i.e. sequences that shift pulse phases on shorter timsecales. For the sequences derived here, control bandwidth is independent of sequence size, so the effective bandwidth of fidelity saturates. Control bandwidth can be increased by choice of base pulse; both composite pulse types used here require larger control bandwidth than a simple pulse, and thus produce significantly better performance.\\
\begin{figure}[!h]
 \includegraphics[width=87mm]{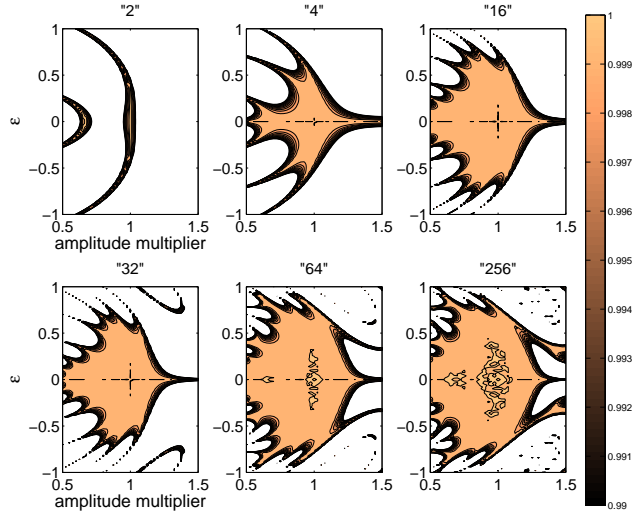}
\caption{\label{fig:fig3} Robustness of sequences versus systematic amplitude errors, for the 3-pulse composite base pulse. Sequence fidelity is displayed as a contour map (grayscale), with $f=99\%$ as the cutoff (white areas represent $f<99\%$). The horizontal axes are amplitude scale factors, and the vertical axes are fractional resonance offset $\epsilon$. It can be seen that robustness to amplitude errors generally increases with sequence length, in a similar fashion to offset robustness.}
\end{figure}
\indent Figure~\ref{fig:fig2} shows fidelities versus offset for windowed sequences, i.e. sequences with an added free evolution delay period adjacent to each base pulse. Results are shown for the 3-pulse composite base pulse, for three different delay periods. These free evolution periods cause small-amplitude fidelity oscillations as a function of offset, but do not degrade the overall sequence performance appreciably, regardless of the delay time chosen. This can be seen by comparison with the windowless sequences, figure~\ref{fig:fig1}b. Therefore, any of the sequences can be applied for arbitrary times $T\geq N t_P$ ($t_P$ is the duration of $P$, and $N$ sequence size) with results similar to the corresponding windowless case. \\
\indent Another property of interest for these sequences is the robustness to systematic pulse amplitude error. For zero resonance offset, the sequences are perfectly robust since they correspond to rotations of $0\pi$. Hence, we expect reasonably good amplitude robustness in the range that the pulses continue to compensate well for offset errors. This expectation is borne out in numerical simulations: figure~\ref{fig:fig3} shows sequence fidelity versus both resonance offset and an amplitude scale factor, for several sequences using the 3-pulse composite base pulse. Robustness to amplitude errors increases with sequence size in a similar fashion to offset robustness. For the 3-pulse composite $\pi$, the patterns demonstrate that low-amplitude errors, or under-rotations, are more forgiving than over-rotations. In this case, given a symmetric distribution of amplitudes, the optimal average amplitude for maximizing fidelities over the distribution is $\approx 0.9\nu_1$. However, this behavior is dependent on choice of base pulse; for a simple pulse, the fidelity versus amplitude is more symmetric and centered on scale factor 1. This underrotation property is also shared by the Tycko (7-pulse) composite pulse, and can be used to increase the bandwidth of fidelity at the expense of reducing fidelities for small offsets. Figure~\ref{fig:fig4}a shows the fidelity of the 64-pulse sequence using the Tycko pulse versus offset and amplitude scaling. The vertical dotted line at scale factor $=1$ indicates the cut shown in the figure~\ref{fig:fig1}c, i.e. `correct' amplitude. The left dotted line at scale factor $\approx0.8$ indicates a wider region of offset over which sequence fidelities $f\geq0.999$ are obtained. Hence, by underrotating each base pulse $P$ by $\approx0.8$, a wider bandwidth of refocusing can be achieved in a certain fidelity range. This is somewhat surprising, since lower amplitude corresponds to a smaller bandwidth of excitation. Figure~\ref{fig:fig4}b shows that the same holds true for underrotating by scaling down the timesteps of the base pulse while maintaining the original pulse amplitude. Infact, this method allows even wider bandwidth refocusing: for the 64-pulse sequence, fidelities $f\geq0.999$ are achieved across the range $0\leq|\epsilon|\leq1.2$ when scaling the timesteps by $0.8$. This is can be understood as an increase in the bandwidth of control. The tradeoff is larger infidelity at small offsets, i.e. $f\sim\mathcal{O}(10^{-4})$ for the 64-pulse sequence.
\begin{figure}[!h]
 \includegraphics[width=68mm]{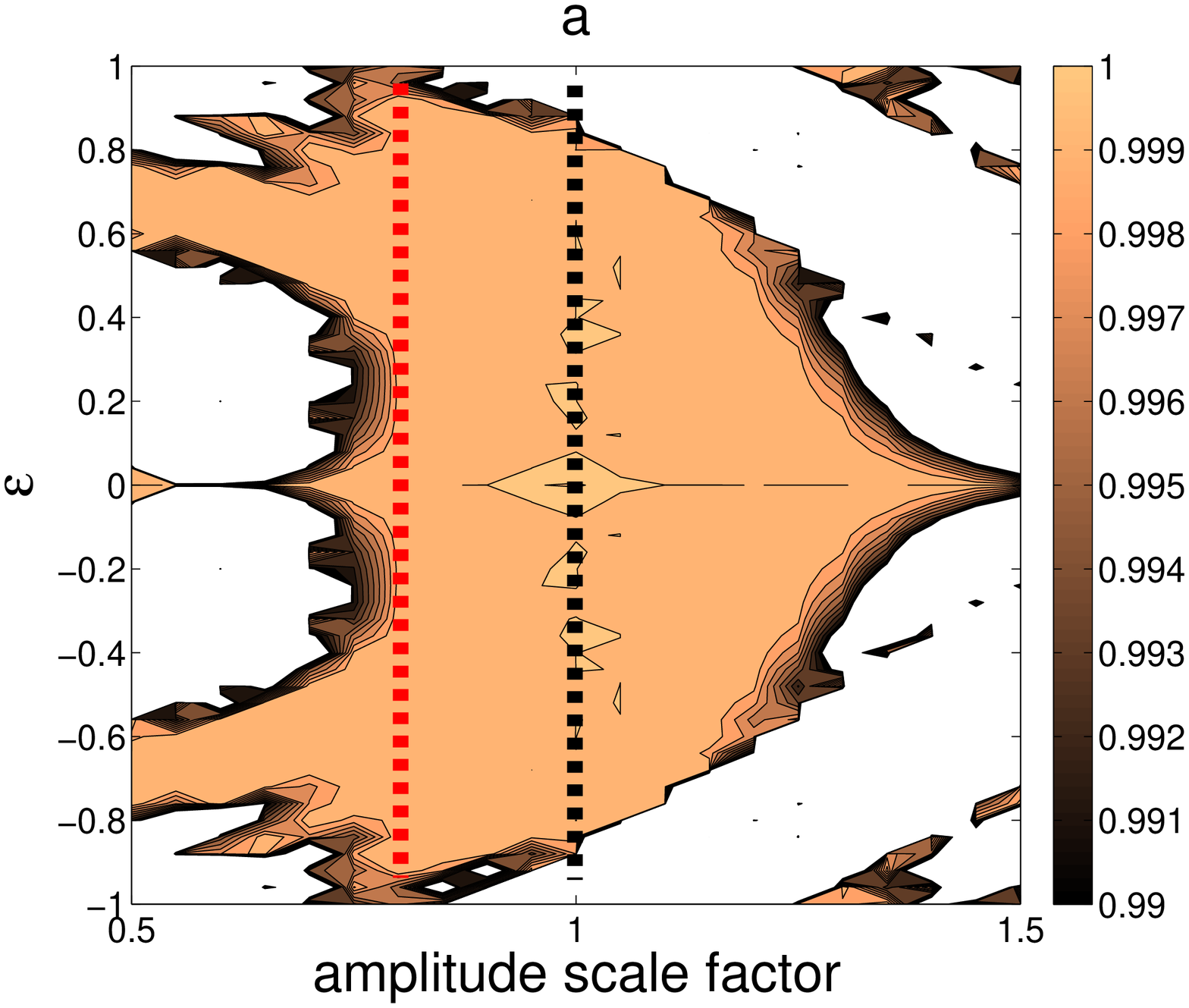}
  \includegraphics[width=65mm]{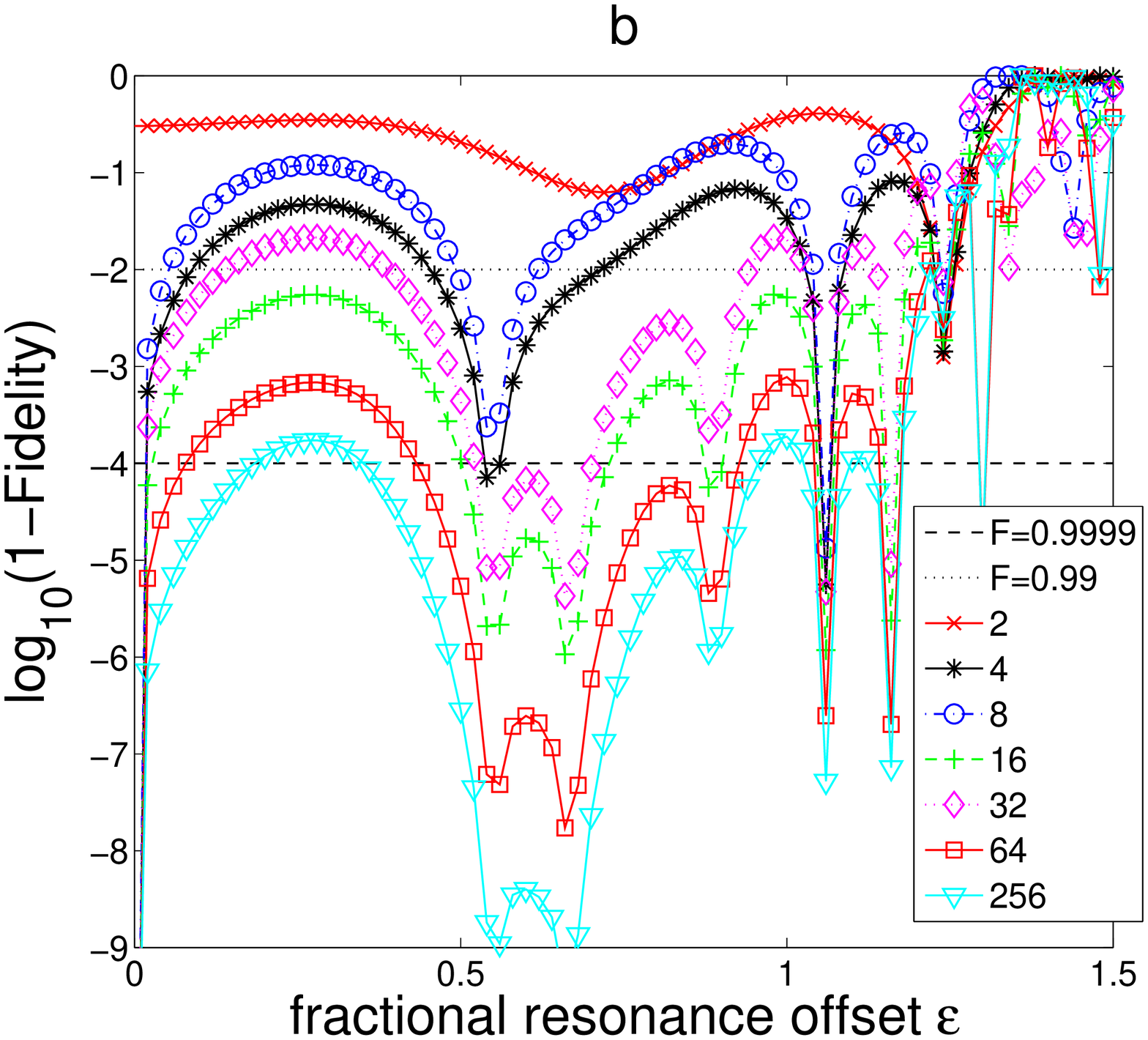}
\caption{\label{fig:fig4} Larger bandwidth refocusing can be achieved for the Tycko (7-pulse) composite base pulse by underrotating, at the expense of reduced fidelities in the low-$\epsilon$ regime. \textbf{(a)} For the 64-pulse sequence, by reducing the pulse amplitude to $\approx 0.8\nu_1$ (left dotted line), the threshold $|\epsilon_{max}|$ for fidelities $f\geq 0.99$ can be pushed closer to $1$. \textbf{(b)} Similarly, by maintaining the original pulse amplitude but rescaling the pulse timesteps by $0.8$, fidelities $f\geq 0.999$ can be achieved to $|\epsilon|\approx1.2$, with the tradeoff of flatter performance for $0\leq|\epsilon|\leq1.2$ compared to the original timestep scaling.}
\end{figure}
\section{Conclusions}
In conclusion, we have shown a new method for constructing error-compensating sequences to approximate the identity operator for the case of resonant field strength comparable to resonance offset. The sequences require only one base pulse, together with its phase reversed twin, lending simplicity for experimental implementation. The construction algorithm makes explicit use of the spinor transformation property as a tool for canceling offset errors. To construct the sequences, we first reduce $\sigma_z$ errors using the spinor rotation property, then reduce transverse error (in proportion to $\sigma_z$ error) by anti-symmetrization, repeating iteratively. The process can, in principle, be carried out to any error order, but the logarithmic elimination of error orders relative to growth of error coefficients limits the gains in refocussing bandwidth for very long sequences. \\
\indent We expect these sequences to be useful for maintaining coherence in any qubit devices for which there exist wide, static distributions of qubit splitting and limited control field amplitude, or for which many measurements must be averaged in the presence of slowly drifting resonance frequencies. One particular example is the presence of random (quasi-static) nuclear fields acting on electron spin qubits in quantum dots; such wideband refocusing with limited controls could provide a useful means to effectively decouple the nuclear system from the qubit. It will be interesting to investigate the efficiency with which slow evolution of the nuclear fields, driven by spectral diffusion, spoils the error cancellation properties of these sequences, i.e. whether such sequences may provide any benefits for decoupling from a dynamic environment. 

\begin{acknowledgments}
We thank C. A. Ryan and W. A. Coish for stimulating discussions, and acknowledge the Natural Sciences and Engineering Research Council for funding. Mathematica and Matlab codes used to perform analytical and numerical calculations are available upon request. 
\end{acknowledgments}
\bibliography{baugh_submit}

\end{document}